\documentclass[12pt]{article}
\usepackage{graphicx}
\usepackage{amssymb,amsmath,bm}
\usepackage{cite}

\newcommand{\lsim}{
\mathrel{\hbox{\rlap{\hbox{\lower4pt\hbox{$\sim$}}}\hbox{$<$}}}}

\newcommand{\gsim}{
\mathrel{\hbox{\rlap{\hbox{\lower4pt\hbox{$\sim$}}}\hbox{$>$}}}}

\newcommand\pubnumber{CERN-PH-TH-2014-240\\ TTP14-033}
\newcommand\pubdate{\today}

\def\address{CERN Theory Division, CH-1211 Geneva 23, Switzerland\vspace{3mm}\\

Institut f\"ur Theoretische Teilchenphysik, Karlsruhe Institute of Technology,\\
Engesserstra{\ss}e 7, D-76128 Karlsruhe, Germany\vspace{3mm}\\

Institut f\"ur Kernphysik, 
Karlsruhe Institute of Technology,\\
Hermann-von-Helmholtz-Platz 1, 76344 Eggenstein-Leopoldshafen}

\def\Title#1{\begin{center} {\Large #1 } \end{center}}
\def\Author#1{\begin{center}{ \sc #1} \end{center}}
\def\Address#1{\begin{center}{ \it #1} \end{center}}

\newcommand\pubblock{\rightline{\begin{tabular}{l} \pubnumber\\
         \pubdate  \end{tabular}}}
\newenvironment{Abstract}{\begin{quotation}  }{\end{quotation}}
\newenvironment{Presented}{\begin{quotation} \begin{center} 
             PRESENTED AT\end{center}\bigskip 
      \begin{center}\begin{large}}{\end{large}\end{center} \end{quotation}}

\def\beq{\begin{equation}}
\def\eeq#1{\label{#1}\end{equation}}
\def\eeqn{\end{equation}}

\def\beqa{\begin{eqnarray}}
\def\eeqa#1{\label{#1}\end{eqnarray}}
\def\eeqan{\end{eqnarray}}

\let\bar=\overbar

\def\Dslash{\not{\hbox{\kern-4pt $D$}}}
\def\dslash{\not{\hbox{\kern-2pt $\del$}}}

\def\msb{{\bar{\ssstyle M \kern -1pt S}}}

\begin{document}
\begin{titlepage}
\pubblock

\vfill
\Title{Flavour Physics Beyond the Standard Model:\\
Recent Developments and Future Perspectives
}
\vfill
\Author{Monika Blanke}
\Address{\address}
\vfill
\begin{Abstract}
Flavour physics plays a crucial role in the search for physics beyond the Standard Model (SM). While $B$ physics offers many observables to look for deviations from the SM, the highest new physics sensitivity can be obtained in the rare kaon decays $K\to\pi\nu\bar\nu$. Of particular interest are correlations between various flavour violating observables that allow to test the symmetries and the operator structure of the new physics flavour sector. New physics searches in rare meson decays are complemented by searches for new flavour violating interactions in the production and decay of new particles at the LHC and in dark matter phenomenology.
\end{Abstract}
\vfill
\begin{Presented}
8th International Workshop on the CKM Unitarity Triangle (CKM 2014),
Vienna, Austria, September 8--12, 2014
\end{Presented}
\vfill
\end{titlepage}
\def\thefootnote{\fnsymbol{footnote}}
\setcounter{footnote}{0}
%



\section{Physics Beyond the Standard Model?}

The Standard Model (SM) has so far been extremely successful in explaining  particle physics data. Nonetheless several reasons let us believe that it is incomplete and needs to be extended by beyond the Standard Model (BSM) particles and interactions. The SM flavour sector introduces a large number of parameters with a very hierarchical pattern. It is expected that a fundamental theory should require a smaller number of parameters and explain the origin of flavour by some (approximate) symmetry. Besides, the SM does not provide a dark matter (DM) candidate and cannot accommodate the dark energy and the baryon asymmetry of the universe. Last but not least within the SM the origin of electroweak (EW) symmetry breaking is left unexplained and the Higgs mass is subject to fine tuning, unless new physics at the TeV scale stabilises it.

Many BSM models have been suggested which solve one or several of these puzzles. Extensive searches have been performed, both directly and indirectly, and high hopes have been pinned especially on the LHC.  Yet unfortunately so far no sign of new physics has been found and the data are in impressive agreement with the SM predictions.

With the 13\,TeV run of the LHC starting soon, we are now confronted with the possibilities that either new particles will be found in the very near future, or that we may have to wait for a very long time in case they are out of the LHC reach. Preparing ourselves for both scenarios, it is crucial to remember the important role of indirect tests of new physics, in particular in the flavour sector.

In the exciting case of a direct LHC discovery, it will be crucial to measure as many observables as possible in order to understand the nature and the coupling structure of the new particle zoo inhabitants. Especially the flavour structure is non-trivial to access in high energy collisions, so that complementary information from the flavour precision frontier is required. On the other hand if no new physics is seen by ATLAS and CMS our best hope will be to look for deviations from the SM in precision data. Again flavour changing neutral current (FCNC) observables play a unique role due to their sensitivity to very high new physics scales, as we discuss next.

\section{New Physics Reach of Flavour Physics}

FCNC processes within the SM suffer from a strong four-fold suppression: Due to the unitarity of the CKM matrix they arise first at the one-loop level, and they are further suppressed by small quark masses, i.\,e.\ the GIM-mechanism \cite{Glashow:1970gm}. Thirdly some FCNC transitions receive a further suppression from the pure $V-A$ structure of charged current interactions mediating flavour violation in the SM. Last but not least any flavour violating effect in the SM is strongly suppressed by the smallness of the CKM mixing angles. The CKM hierarchy also predicts a specific pattern of effects in the various meson systems:
\begin{equation}\label{eq:CKM}
\underbrace{V_{ts}^* V_{td}}_{K\text{ system}} \sim 5\cdot 10^{-4} \ll 
\underbrace{V_{tb}^* V_{td}}_{B_d\text{ system}} \sim  10^{-2} < 
\underbrace{V_{tb}^* V_{ts}}_{B_s\text{ system}} \sim  4\cdot 10^{-2}\,,
\end{equation}
i.\,e.\ FCNC and CP violating rates are generally smallest in the kaon system.

In principle all of these suppression mechanisms can be absent in BSM scenarios, so that large deviations from the SM predictions can generally be expected in FCNC observables, especially in the $K$ meson sector. However such large deviations have not been observed, telling us that either new physics must involve some kind of suppression of FCNC effects, or that it can arise only at very high scales. 

The latter leads us to the question which scales can possibly be probed by rare flavour violating decays.
The pattern of CKM suppressions \eqref{eq:CKM} lets us suspect that rare kaon decays be the best place to look for new physics at very high energy scales. Particularly suited are the decays $K^+\to\pi^+\nu\bar\nu$ and $K_L\to\pi^0\nu\bar\nu$ which are famous for their outstanding theoretical cleanliness and extreme suppression within the SM. 

\begin{figure}[htb]
\centering
\includegraphics[width=.49\textwidth]{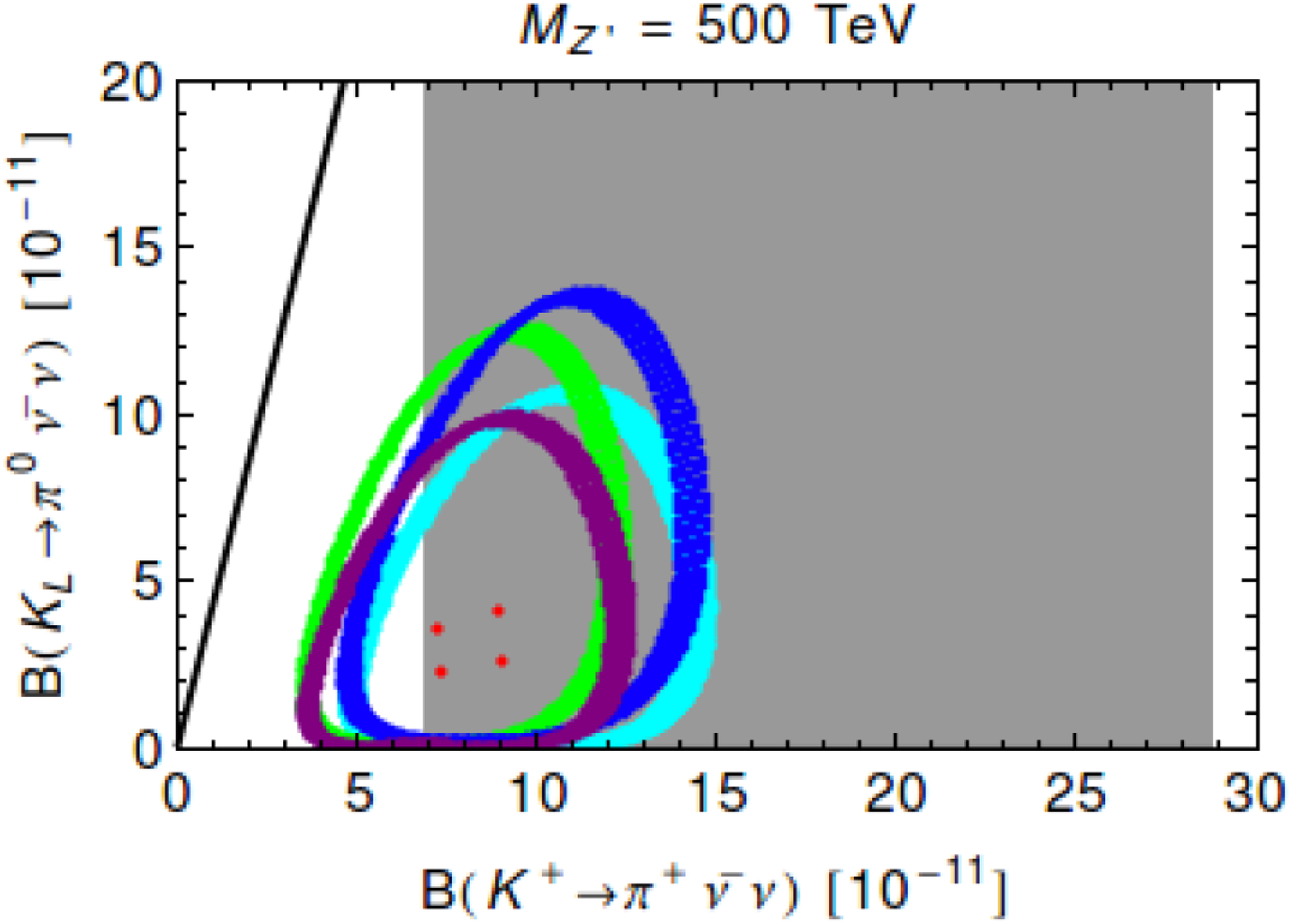}\hfill
\includegraphics[width=.49\textwidth]{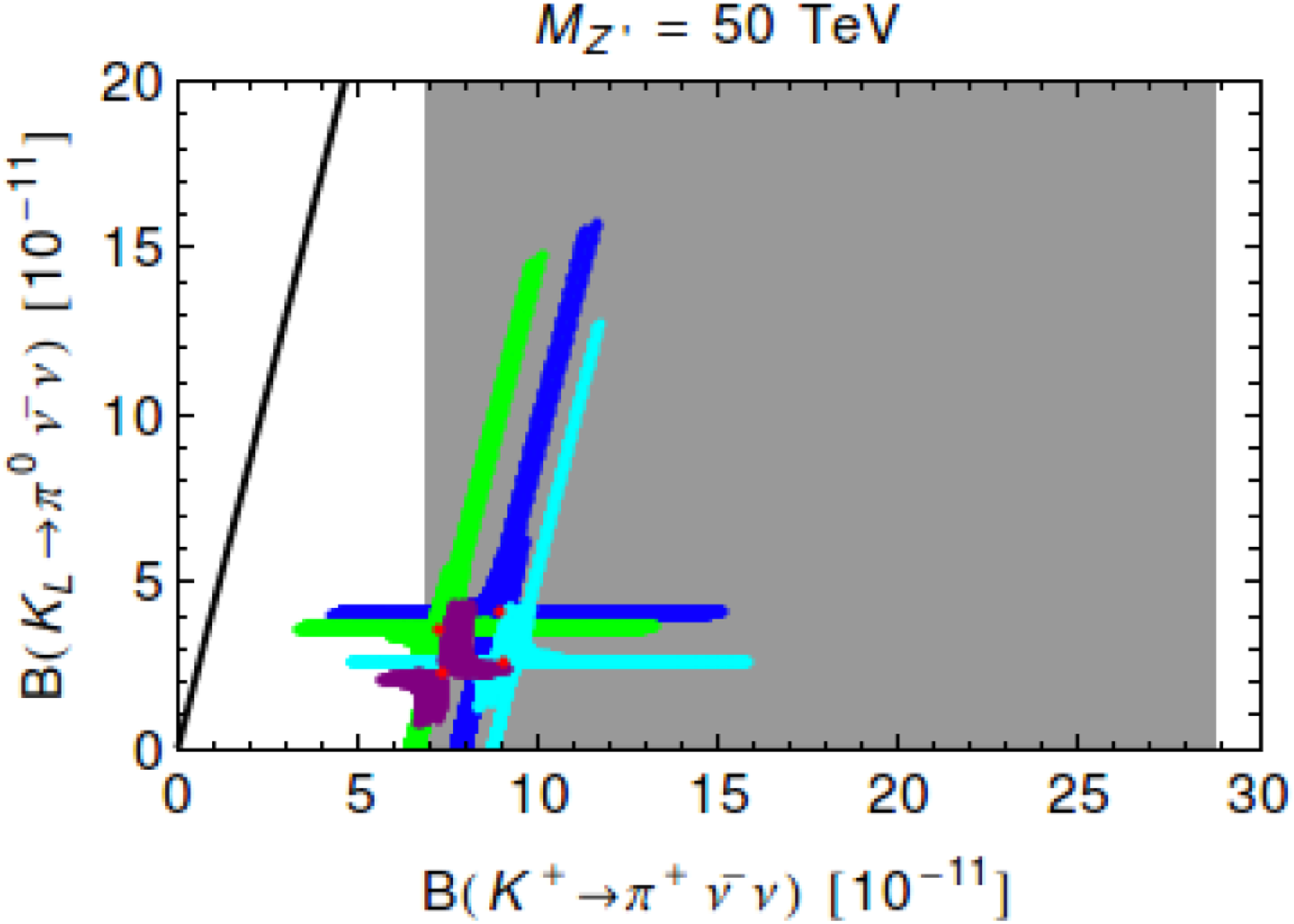}
\caption{The charged and neutral $K\to\pi\nu\bar\nu$ branching ratios for various realisations of a flavour changing $Z'$ gauge boson \cite{Buras:2014zga}. }
\label{fig:zepto}
\end{figure}

Indeed in \cite{Buras:2014zga}, studying the effects of a flavour violating $Z'$ gauge boson, these channels have been found to be sensitive to new physics scales as large as $\sim 1000\,\text{TeV}$ even after taking into account the constraints from $K^0-\bar K^0$ mixing. The $K\to\pi\nu\bar\nu$ decays are thus sensitive to the {\it zeptouniverse}, corresponding to length scales $\lsim 10^{-21}\,\text{m}$ or equivalently $\gsim 200\,\text{TeV}$. The left plot in figure \ref{fig:zepto} shows the effects of a flavour changing $Z'$ with mass $M_{Z'} = 500\,\text{TeV}$ and both left- and right-handed couplings to quarks. The sizeable deviations from the SM value are possible thanks to the possible cancellation of the various operator contributions to $K^0-\bar K^0$ mixing. If on the other hand, as shown in the right plot, only left- (or right-)handed couplings to quarks are present, the new physics reach of the $K\to\pi\nu\bar\nu$ system is somewhat more modest -- yet still better than that of any direct search experiment in the foreseeable future.

While rare $K$ decays are in general most sensitive to very high new physics scales, we note in passing that also the $B_{s,d}\to\mu^+\mu^-$ decays are able to probe the zeptouniverse \cite{Buras:2014zga}, at least in the case of a heavy flavour violating neutral scalar.

Before moving on, let us again turn our attention to the plots in figure \ref{fig:zepto}. Apart from the vastly different $Z'$ mass scales, there are two other striking observations to be made. 

First, the predictions for the $K\to\pi\nu\bar\nu$ branching ratios depend strongly on the values of the CKM parameters $|V_{ub}|$ and $|V_{cb}|$, whose inclusive and exclusive values are indicated by the different colours in the plots. A precise determination of these parameters is therefore crucial to fully explore the potential of rare decays.

Second, the presence of only left- (or right-)handed $Z'$ interactions leads to a very specific cross-like correlation in the $K\to\pi\nu\bar\nu$ plane. This correlation has been observed in other models with only left-handed flavour violating interactions before (see e.\,g.\ \cite{Blanke:2006eb,Promberger:2007py,Buras:2012jb}) and has been studied in a model independent manner in \cite{Blanke:2009pq}. The observed cross structure is a direct consequence of the strong constraint from CP violation in $K^0-\bar K^0$ mixing, measured by the parameter $\varepsilon_K$, allowing only for very specific phases of the new physics $s\to d$ amplitude. Such direct correspondence between $\Delta S=2$ and $\Delta S=1$ processes is however possible only in the absence of the chirally enhanced left-right operator contributions to $K^0-\bar K^0$ mixing. In this sense the correlation between the charged and neutral $K\to\pi\nu\bar\nu$ modes can be viewed as a test of the operator structure in $K^0-\bar K^0$ mixing.

\section{Rare Decays and their Correlations}

As we have just seen for the example of $K\to\pi\nu\bar\nu$ decays, correlations between FCNC observables play a crucial role in deciphering the origin of new flavour violating effects. With the ultimate goal  to understand the underlying symmetries and coupling structure of the new physics at work, it is therefore mandatory to obtain as many precise measurements of flavour violating observables as possible and to look for model distinguishing correlations between them. 

Generally such correlations can be divided into two classes. Correlations between different observables within a given meson system are sensitive to the new physics operator structure, as we have seen in the case of $K\to\pi\nu\bar\nu$. Correlations between different meson systems on the other hand test the possible  flavour symmetries of the model at work. For example models with a $U(2)^3$ flavour symmetry predict specific relations between $b\to d$ and $b\to s$ transitions.

\begin{figure}[htb]
\centering
\includegraphics[width=.47\textwidth]{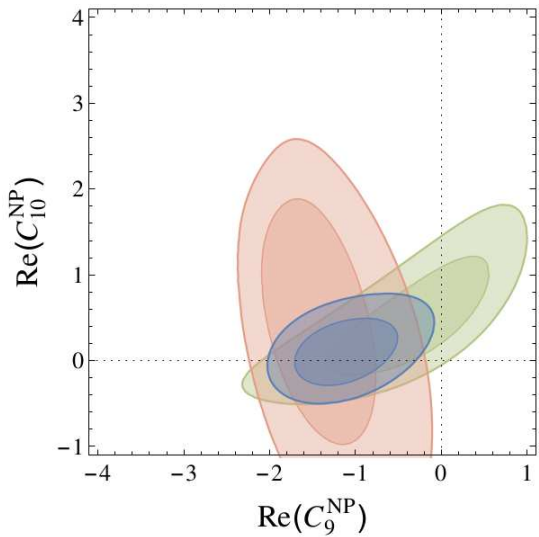}
\caption{Model independent fit for the Wilson coefficients $C_9$ and $C_{10}$ \cite{Altmannshofer:2014rta}.}
\label{fig:C9C10}
\end{figure}

A prime example for testing the new physics operator structure is given by the model independent studies of radiative and semileptonic $b\to s$ transitions, like $B\to X_s\gamma$, $B\to K^{(*)}\mu^+\mu^-$ etc. After LHCb announced a $3.7\sigma$ local discrepancy in $P'_5$ \cite{Aaij:2013qta}, one of the angular observables describing the $B\to K^*\mu^+\mu^-$ differential decay rate, this field received particular theoretical attention. Several model independent fits to the Wilson coefficients of the $b\to s\mu^+\mu^-$ effective Hamiltonian have been performed \cite{Descotes-Genon:2013wba,Altmannshofer:2013foa,Beaujean:2013soa,Altmannshofer:2014rta}, with the outcome that a large negative contribution $C_9^\text{NP}$ to the effective four fermion coupling $(\bar bs)_{V-A}(\bar\mu\mu)_V$ is necessary to accommodate the observed deviation. The most recent fit can be found in \cite{Altmannshofer:2014rta}, see also figure \ref{fig:C9C10}.

While such large deviation from the SM, if confirmed, would be an exciting new physics signature it turns out to be rather difficult to generate in popular extensions of the SM. Both in supersymmetric models and in models with partial compositeness, $b\to s\mu^+\mu^-$ transitions are dominated by $Z$ boson exchanges. The vectorial coupling of the $Z$ boson to muons is however accidentally suppressed so that it appears impossible to generate a large shift in $C_9$ without affecting $C_{10}$ in an even more pronounced manner.
An interesting model which can explain the LHCb data has been proposed in \cite{Altmannshofer:2014cfa}. It is based on gauging $L_\mu - L_\tau$ number, where $L_i$ denotes the lepton flavour number of $i=\mu,\tau$.

\begin{figure}[htb]
\centering
\includegraphics[width=.6\textwidth]{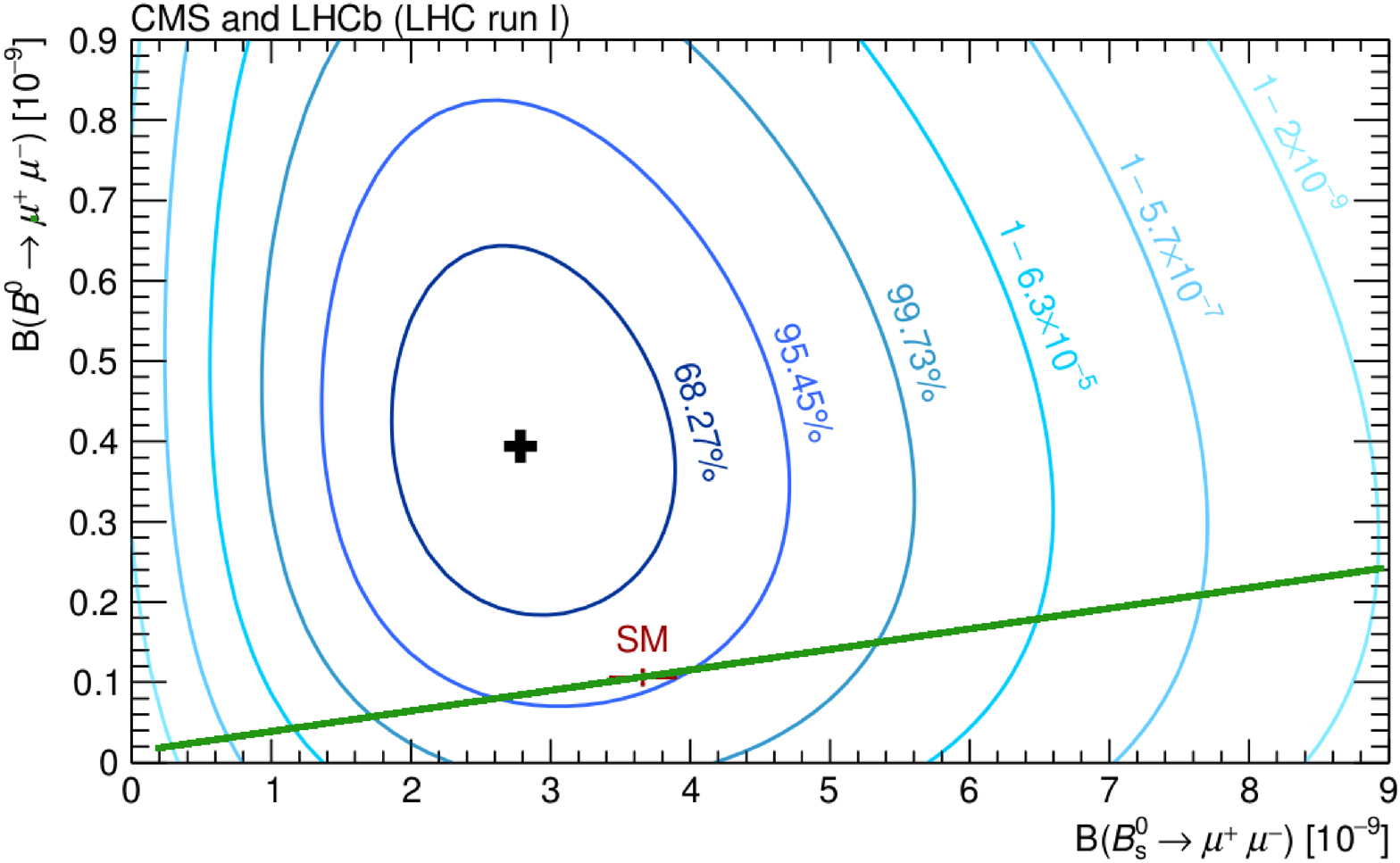}
\caption{Branching ratios of $B_{s,d}\to \mu^+\mu^-$ \cite{CMS:2014xfa}.}
\label{fig:Bsdmumu}
\end{figure}

Another correlation that has recently attracted a lot of attention is the one between the branching ratios  $\bar{\mathcal B}(B_s\to\mu^+\mu^-)$ and $\mathcal{B}(B_d\to\mu^+\mu^-)$. With the recent measurement of the flavour averaged $B_s\to\mu^+\mu^-$ branching ratio by LHCb and CMS \cite{Aaij:2013aka,Chatrchyan:2013bka,CMS:2014xfa} and the two-sided bound on $B_d\to\mu^+\mu^-$, we have at hand not only yet another important probe of the SM, but -- in case of a deviation from the SM prediction -- at the same time an equally powerful test of the new physics flavour structure. Any deviation from the straight green line indicated in figure \ref{fig:Bsdmumu} would rule out the Minimal Flavour Violation hypothesis \cite{Buras:2000dm,D'Ambrosio:2002ex,Hurth:2008jc} as well as models with a $U(2)^3$ flavour symmetry \cite{Barbieri:2011ci,Barbieri:2012uh,Buras:2012sd}. While due to the large uncertainties the experimental results are currently fully consistent with the SM prediction, there is still a lot of room for a striking deviation on which the 13\,TeV LHC run may shed light.

To conclude this section, let us stress once more the importance of rare flavour and CP violating decays in searching for physics beyond the SM. Rare meson decays have been an active and successful field of research for many decades. They offer a plethora of observables, many of which by now have reached an impressive precision. As we have seen correlations between different decay modes play a central role in constraining the SM and new physics flavour structure.

\section{Interplay with other New Physics Searches}

Despite the great success of $K$ and $B$ physics in constraining the SM, unfortunately so far no clear sign of new physics has been found in these fields. Furthermore even if one or several of the present small hints for a deviation will eventually become a convincing discrepancy, the precesses at hand only provide an indirect probe of new flavour violating interactions. Additionally with rare $K$ and $B$ decays only it is difficult, if not impossible, to access flavour violation in top and Higgs couplings. Therefore in order to fully exploit the new physics flavour structure new complementary observables are needed. Besides the direct searches for flavour violating top and Higgs couplings and charged lepton flavour violating processes, these include the study of flavour violating interactions of new particles at the LHC and the phenomenology of flavoured dark matter. In this section we will briefly review an example for each of the latter two. We start with the LHC phenomenology of flavour violating stops, the supersymmetric partners of the top quark. Subsequently we turn our attention to a simplified model for dark matter carrying flavour quantum number and mediating flavour violation to the SM quark sector.

\subsection{\boldmath Stop flavour violation at the LHC}

Models that address the naturalness problem of electroweak symmetry breaking typically introduce new coloured particles below the TeV scale, the so-called top partners. They cancel the quadratic divergence to the Higgs mass arising from top quark loops and therefore couple to the Higgs boson via the Yukawa coupling $Y_t$. In general however these top partners need not be mass eigenstates, so that their production and decay will lead to flavour violating signatures at the LHC, often with up or charm quarks in the final state.

\begin{figure}[htb]
\centering
\includegraphics[width=.55\textwidth]{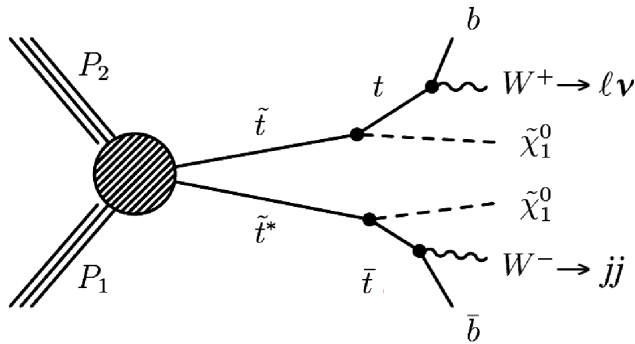}
\caption{Top squark pair production at the LHC.}
\label{fig:stop-pair-prod}
\end{figure}

In supersymmetry the top squarks, also called stops, are pair produced via strong interactions. Their decay is model dependent, however searches usually assume an $\mathcal{O}(1)$ branching ratio into $t+\text{LSP}$, where the lightest supersymmetric particle (LSP) is assumed to be stable and escapes detection. The experimental signature then depends on the decays of the final state $t\bar t$ pair, see figure \ref{fig:stop-pair-prod} for an example. Despite extensive searches at the LHC experiments, no sign of stops has yet been seen and the bounds on their masses become increasingly strong, undermining the natural motivation.

\begin{figure}[htb]
\centering
\includegraphics[width=.5\textwidth]{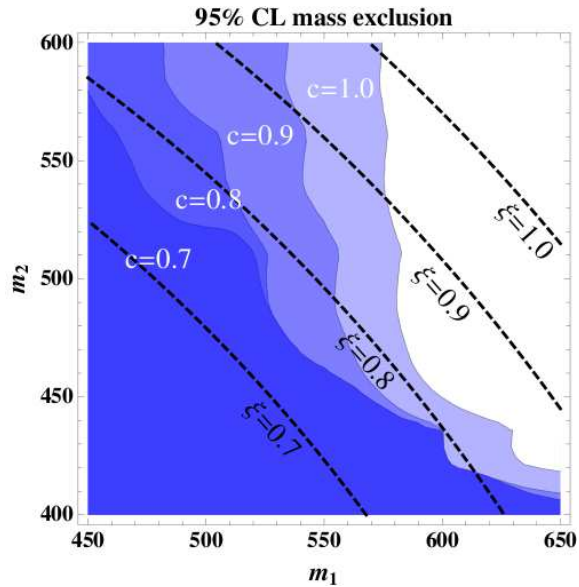}
\caption{Mass bounds in the mixed $\tilde t_R - \tilde c_R$ system \cite{Blanke:2013uia}.}
\label{fig:stop-scharm-mass}
\end{figure}

These bounds however become invalid once flavour violation is taken into account. To see this, let us consider the simple scenario where the right-handed stop $\tilde t_R$ mixes with the right-handed scharm $\tilde c_R$, as done in \cite{Blanke:2013uia}.\footnote{For earlier related studies, see e.\,g.\ \cite{Han:2003qe,
Cao:2006xb,
Hiller:2008wp,
Kribs:2009zy,
Hurth:2009ke,
Bruhnke:2010rh,
Bartl:2010du,Bartl:2012tx}.} If the stop scharm mixing angle $c = \cos\theta$ deviates significantly from $c=1$, the flavour conserving case, then the branching ratio of the stop-like state $\tilde q_1 \to t+\text{LSP}$ receives a suppression factor $c^2$. At the same time the decay channel $\tilde q_1 \to c+\text{LSP}$ opens up, leading to a light jet signature which is much less constrained at the LHC; and vice versa for the charm-like squark $\tilde q_2$. Taking into account simultaneously the production and decay of $\tilde q_1$ and $\tilde q_2$, the modified constraints have been estimated in \cite{Blanke:2013uia}, using the experimental searches for stops \cite{ATLAS:2012maq} and light squarks \cite{ATLAS:2012ona,Aad:2012fqa,Chatrchyan:2012lia,CMS:2012hqo,Chatrchyan:2012wa,Mahbubani:2012qq}.
The resulting $2d$ exclusion contours are shown in figure \ref{fig:stop-scharm-mass} for different values of the mixing angle $c$. A large flavour mixing angle allows for a significant reduction in the allowed stop-like mass, and therefore in a lower rate of fine tuning $\xi<1$.

While the analysis in  \cite{Blanke:2013uia} dealt with the case of stops in the several hundred GeV range, flavour mixing also has interesting implications for stops below the $m_t + m_{\chi_1^0}$ threshold. This case has been studied in detail in \cite{Agrawal:2013kha}. In this setup the stop decays dominantly into a light jet and the LSP in a large region of parameter space. The NLO corrections to the decay in question have been evaluated in \cite{Grober:2014aha,Aebischer:2014lfa} and found to be sizeable. A dedicated search by the ATLAS collaboration \cite{TheATLAScollaboration:2013aia} was able to set constraints of up to $230\,\text{GeV}$ on the stop mass.

\subsection{Flavoured Dark Matter}

The existence of dark matter (DM) is well-established by cosmological and astrophysical observations, yet so far no direct evidence of its particle nature has been obtained. This leaves a lot of room for speculations on the quantum numbers and interactions of DM. Numerous models have been suggested and studied. A class of models that recently received particular attention are those with {\it flavoured DM}, i.\,e.\ DM carries flavour charge and has flavour violating interactions to quarks or leptons (see e.\,g.\ \cite{Kile:2011mn,Batell:2011tc,Kamenik:2011nb,Agrawal:2011ze,Kumar:2013hfa,Kile:2013ola,Batell:2013zwa
}). 

\begin{figure}[htb]
\centering
\includegraphics[width=.36\textwidth]{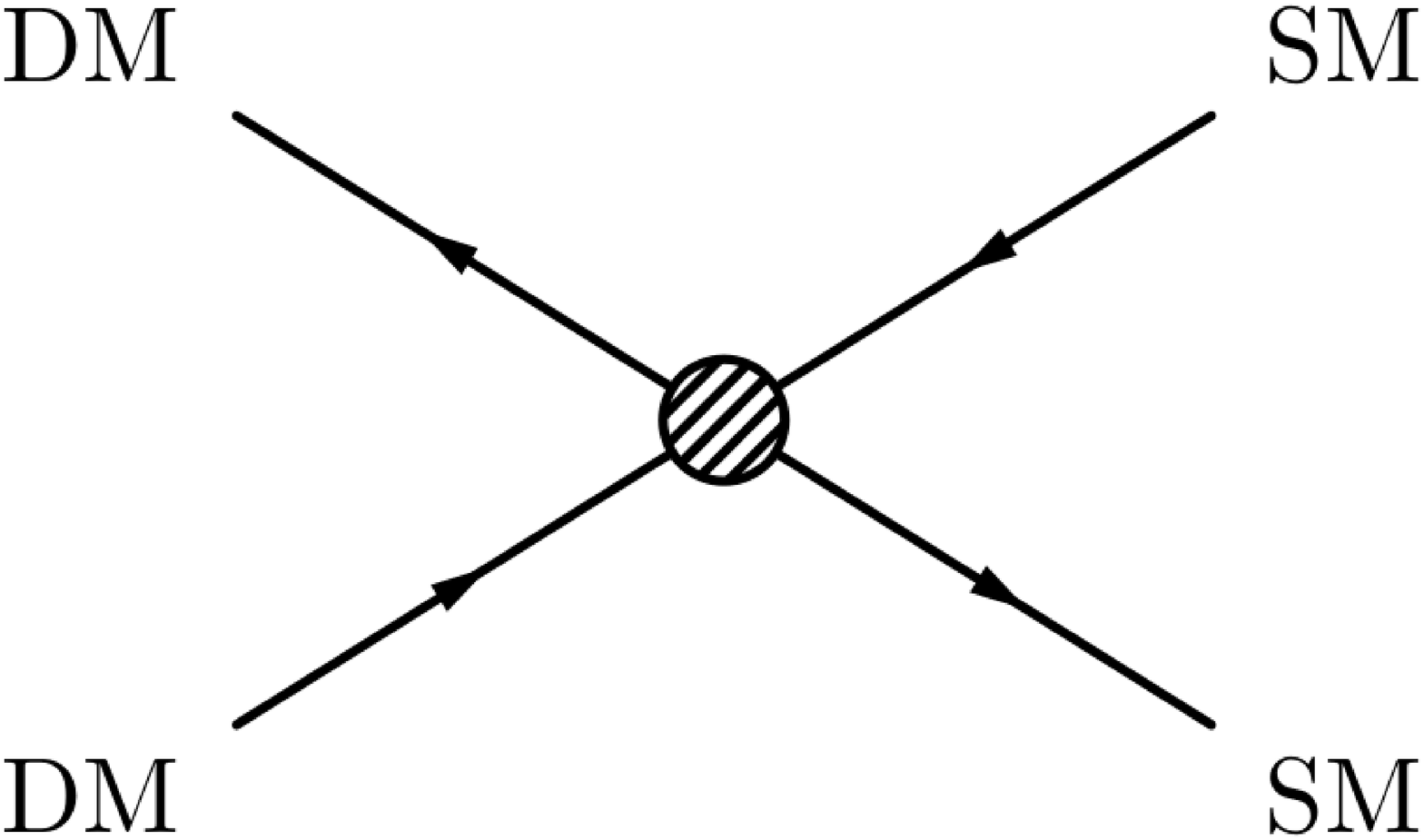}\hspace{1cm}
\includegraphics[width=.35\textwidth]{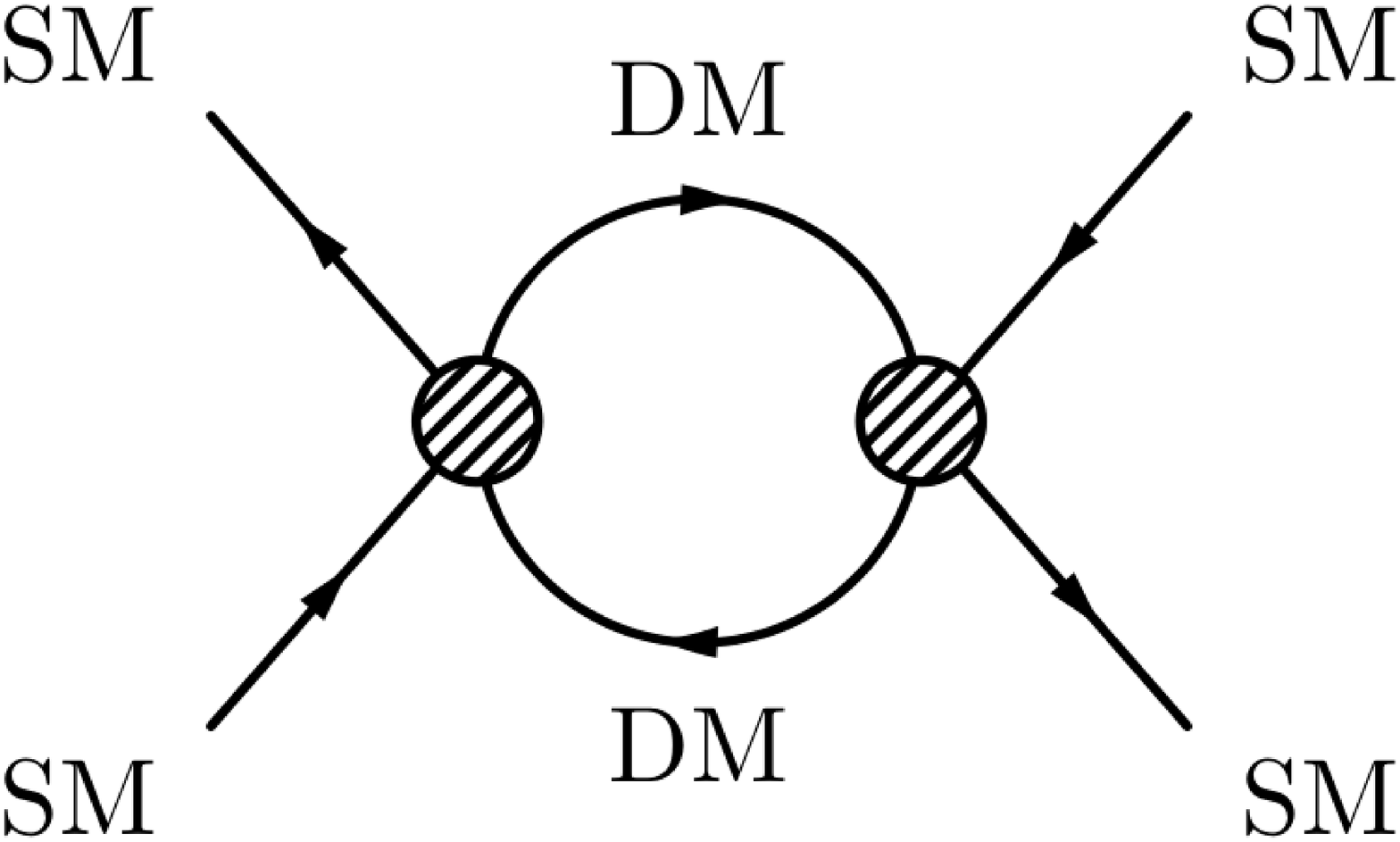}
\caption{Effective tree and one-loop diagrams relevant for the phenomenology of flavoured dark matter \cite{Agrawal:2014aoa}.}
\label{fig:dm-search}
\end{figure}

Flavoured DM as usual generates an effective four point interaction between DM and SM particles as shown in the left diagram in figure \ref{fig:dm-search}, giving rise to possible signatures in direct and indirect detection experiments and at the LHC. In addition the flavour violating structure of the interaction also leads to flavour changing processes with the new particles contributing through loops, as shown in the right diagram in figure \ref{fig:dm-search}.

Most studies so far focused on Minimal Flavour Violation scenarios in order to avoid the stringent flavour physics constraints. Recently however a viable scenario for non-minimally flavour violating dark matter has been proposed \cite{Agrawal:2014aoa}. In this simplified model DM is introduced as a flavoured Dirac fermion $\chi$ that couples to right handed down type quarks via a scalar mediator. The coupling matrix $\lambda$ is assumed to be the only new source of flavour violation, based on which this setup is named {\it Dark Minimal Flavour Violation}. 

The flavour, dark matter, and collider phenomenology has also been studied in \cite{Agrawal:2014aoa}. $\Delta F =2$ constraints restrict $\lambda$ to a very non-generic structure, with either quasi-degenerate diagonal entries or small mixing angles. The new contributions to $\Delta F =1$ rare decays, as well as to electroweak precision observables and electric dipole moments have been found to be small. Collider constraints can be recast from SUSY searches at the LHC, with strong constraints up to $850\,\text{GeV}$ emerging on the mass of the scalar mediator. 

\begin{figure}[htb]
\centering
\includegraphics[width=.6\textwidth]{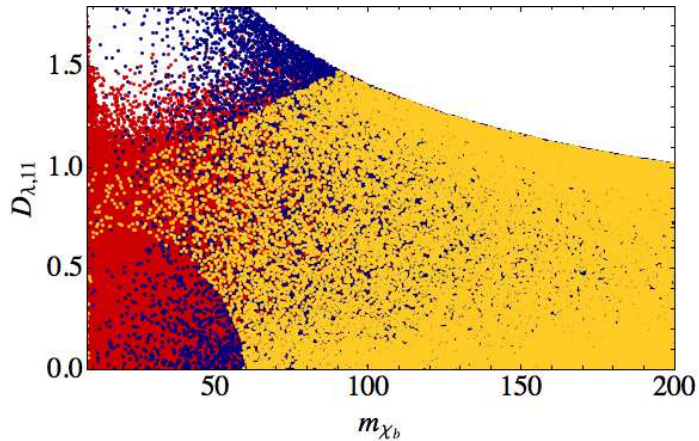}
\caption{Allowed range for the coupling $D_{\lambda,11}$ as a function of the dark matter mass $m_{\chi_b}$ \cite{Agrawal:2014aoa}.}
\label{fig:mchi-D11}
\end{figure}

A particularly interesting interplay of constraints is found when considering simultaneously the constraints from FCNC observables and from direct DM detection, as shown in figure \ref{fig:mchi-D11}. While the flavour (blue) and DM data (red) separately do not constrain the size of the first quark generation coupling $D_{\lambda,11}$ to the new sector, the interplay of both constraints (yellow) yields both an upper and a lower bound for small values of the DM mass.

\section{Conclusions}

Irrespective of the actual discoveries during the next LHC run, flavour physics is and will remain a powerful probe of BSM interactions. On the one hand rare kaon decays, in particular the clean channels $K\to\pi\nu\bar\nu$, have the highest discovery reach, with sensitivity beyond $10^3\,\text{TeV}$. On the other hand $B$ meson decays offer a very large number of observables that are well suited to  test the SM flavour structure.
In this context it is of utmost importance to study correlations between observables, as those are crucial to pin down the new physics flavour structure. Last but not least the interplay with other new physics searches (like the LHC, dark matter searches etc.) provides a useful complementary tool towards understanding the theory of flavour that certainly needs further investigation.


\begin{thebibliography}{999}

\bibitem{Glashow:1970gm}
  S.~L.~Glashow, J.~Iliopoulos and L.~Maiani,
  Phys.\ Rev.\ D {\bf 2} (1970) 1285.


\bibitem{Buras:2014zga}
  A.~J.~Buras, D.~Buttazzo, J.~Girrbach-Noe and R.~Knegjens,
  arXiv:1408.0728 [hep-ph].

\bibitem{Blanke:2006eb}
  M.~Blanke, A.~J.~Buras, A.~Poschenrieder, S.~Recksiegel, C.~Tarantino, S.~Uhlig and A.~Weiler,
  JHEP {\bf 0701} (2007) 066
  [hep-ph/0610298].

\bibitem{Promberger:2007py}
  C.~Promberger, S.~Schatt and F.~Schwab,
  Phys.\ Rev.\ D {\bf 75} (2007) 115007
  [hep-ph/0702169 [HEP-PH]].

\bibitem{Buras:2012jb}
  A.~J.~Buras, F.~De Fazio and J.~Girrbach,
  JHEP {\bf 1302} (2013) 116
  [arXiv:1211.1896 [hep-ph]].


\bibitem{Blanke:2009pq}
  M.~Blanke,
  Acta Phys.\ Polon.\ B {\bf 41} (2010) 127
  [arXiv:0904.2528 [hep-ph]].

\bibitem{Aaij:2013qta}
  R.~Aaij {\it et al.}  [LHCb Collaboration],
  Phys.\ Rev.\ Lett.\  {\bf 111} (2013) 19,  191801
  [arXiv:1308.1707 [hep-ex]].


\bibitem{Descotes-Genon:2013wba}
  S.~Descotes-Genon, J.~Matias and J.~Virto,
  Phys.\ Rev.\ D {\bf 88} (2013) 7,  074002
  [arXiv:1307.5683 [hep-ph]].


\bibitem{Altmannshofer:2013foa}
  W.~Altmannshofer and D.~M.~Straub,
  Eur.\ Phys.\ J.\ C {\bf 73} (2013) 2646
  [arXiv:1308.1501 [hep-ph]].

\bibitem{Beaujean:2013soa}
  F.~Beaujean, C.~Bobeth and D.~van Dyk,
  Eur.\ Phys.\ J.\ C {\bf 74} (2014) 2897
  [arXiv:1310.2478 [hep-ph]].



\bibitem{Altmannshofer:2014rta}
  W.~Altmannshofer and D.~M.~Straub,
  arXiv:1411.3161 [hep-ph].

\bibitem{Altmannshofer:2014cfa}
  W.~Altmannshofer, S.~Gori, M.~Pospelov and I.~Yavin,
  Phys.\ Rev.\ D {\bf 89} (2014) 095033
  [arXiv:1403.1269 [hep-ph]].

\bibitem{Aaij:2013aka}
  R.~Aaij {\it et al.}  [LHCb Collaboration],
  Phys.\ Rev.\ Lett.\  {\bf 111} (2013) 101805
  [arXiv:1307.5024 [hep-ex]].

\bibitem{Chatrchyan:2013bka}
  S.~Chatrchyan {\it et al.}  [CMS Collaboration],
  Phys.\ Rev.\ Lett.\  {\bf 111} (2013) 101804
  [arXiv:1307.5025 [hep-ex]].



\bibitem{CMS:2014xfa}
  V.~Khachatryan {\it et al.}  [CMS and LHCb Collaborations],
  arXiv:1411.4413 [hep-ex].

\bibitem{Buras:2000dm}
  A.~J.~Buras, P.~Gambino, M.~Gorbahn, S.~Jager and L.~Silvestrini,
  Phys.\ Lett.\ B {\bf 500} (2001) 161
  [hep-ph/0007085].

\bibitem{D'Ambrosio:2002ex}
  G.~D'Ambrosio, G.~F.~Giudice, G.~Isidori and A.~Strumia,
  Nucl.\ Phys.\ B {\bf 645} (2002) 155
  [hep-ph/0207036].

\bibitem{Hurth:2008jc}
  T.~Hurth, G.~Isidori, J.~F.~Kamenik and F.~Mescia,
  Nucl.\ Phys.\ B {\bf 808} (2009) 326
  [arXiv:0807.5039 [hep-ph]].

\bibitem{Barbieri:2011ci}
  R.~Barbieri, G.~Isidori, J.~Jones-Perez, P.~Lodone and D.~M.~Straub,
  Eur.\ Phys.\ J.\ C {\bf 71} (2011) 1725
  [arXiv:1105.2296 [hep-ph]].

\bibitem{Barbieri:2012uh}
  R.~Barbieri, D.~Buttazzo, F.~Sala and D.~M.~Straub,
  JHEP {\bf 1207} (2012) 181
  [arXiv:1203.4218 [hep-ph]].


\bibitem{Buras:2012sd}
  A.~J.~Buras and J.~Girrbach,
  JHEP {\bf 1301} (2013) 007
  [arXiv:1206.3878 [hep-ph]].



\bibitem{Blanke:2013uia}
  M.~Blanke, G.~F.~Giudice, P.~Paradisi, G.~Perez and J.~Zupan,
  JHEP {\bf 1306} (2013) 022
  [arXiv:1302.7232 [hep-ph]].



\bibitem{Han:2003qe} 
  T.~Han, K.~-i.~Hikasa, J.~M.~Yang and X.~-m.~Zhang,
  Phys.\ Rev.\ D {\bf 70}, 055001 (2004)
  [hep-ph/0312129].



\bibitem{Cao:2006xb}
J.~Cao, G.~Eilam, K.~-i.~Hikasa and J.~M.~Yang,
Phys.\ Rev.\ D {\bf 74} (2006) 031701
[hep-ph/0604163].



\bibitem{Hiller:2008wp} 
  G.~Hiller and Y.~Nir,
  JHEP {\bf 0803}, 046 (2008)
  [arXiv:0802.0916 [hep-ph]].

\bibitem{Kribs:2009zy} 
  G.~D.~Kribs, A.~Martin and T.~S.~Roy,
  JHEP {\bf 0906}, 042 (2009)
  [arXiv:0901.4105 [hep-ph]].

\bibitem{Hurth:2009ke} 
  T.~Hurth and W.~Porod,
  JHEP {\bf 0908}, 087 (2009)
  [arXiv:0904.4574 [hep-ph]].
  

  

\bibitem{Bruhnke:2010rh} 
  M.~Bruhnke, B.~Herrmann and W.~Porod,
  JHEP {\bf 1009}, 006 (2010)
  [arXiv:1007.2100 [hep-ph]].

\bibitem{Bartl:2010du} 
  A.~Bartl, H.~Eberl, B.~Herrmann, K.~Hidaka, W.~Majerotto and W.~Porod,
  Phys.\ Lett.\ B {\bf 698}, 380 (2011)
  [Erratum-ibid.\ B {\bf 700}, 390 (2011)]
  [arXiv:1007.5483 [hep-ph]].

\bibitem{Bartl:2012tx} 
  A.~Bartl, H.~Eberl, E.~Ginina, B.~Herrmann, K.~Hidaka, W.~Majerotto and W.~Porod,
  arXiv:1212.4688 [hep-ph].
  


\bibitem{ATLAS:2012maq}
  [ATLAS Collaboration],
  ATLAS-CONF-2012-166, ATLAS-COM-CONF-2012-200.


\bibitem{ATLAS:2012ona} 
  [ATLAS Collaboration],
  ATLAS-CONF-2012-109, ATLAS-COM-CONF-2012-140.


\bibitem{Aad:2012fqa}
  G.~Aad {\it et al.}  [ATLAS Collaboration],
  Phys.\ Rev.\ D {\bf 87} (2013) 012008
  [arXiv:1208.0949 [hep-ex]].

\bibitem{Chatrchyan:2012lia}
  S.~Chatrchyan {\it et al.}  [CMS Collaboration],
  Phys.\ Rev.\ Lett.\  {\bf 109} (2012) 171803
  [arXiv:1207.1898 [hep-ex]].

\bibitem{CMS:2012hqo}
  CMS Collaboration [CMS Collaboration],
  CMS-PAS-SUS-12-005.

\bibitem{Chatrchyan:2012wa}
  S.~Chatrchyan {\it et al.}  [CMS Collaboration],
  JHEP {\bf 1301} (2013) 077
  [arXiv:1210.8115 [hep-ex]].

\bibitem{Mahbubani:2012qq}
  R.~Mahbubani, M.~Papucci, G.~Perez, J.~T.~Ruderman and A.~Weiler,
  Phys.\ Rev.\ Lett.\  {\bf 110} (2013) 15,  151804
  [arXiv:1212.3328 [hep-ph]].

\bibitem{Agrawal:2013kha}
  P.~Agrawal and C.~Frugiuele,
  JHEP {\bf 1401} (2014) 115
  [arXiv:1304.3068 [hep-ph], arXiv:1304.3068].

\bibitem{Grober:2014aha}
  R.~Grober, M.~Muhlleitner, E.~Popenda and A.~Wlotzka,
  arXiv:1408.4662 [hep-ph].

\bibitem{Aebischer:2014lfa}
  J.~Aebischer, A.~Crivellin and C.~Greub,
  arXiv:1410.8459 [hep-ph].

\bibitem{TheATLAScollaboration:2013aia}
  The ATLAS collaboration,
  ATLAS-CONF-2013-068, ATLAS-COM-CONF-2013-076.


\bibitem{Kile:2011mn}
  J.~Kile and A.~Soni,
  Phys.\ Rev.\ D {\bf 84} (2011) 035016
  [arXiv:1104.5239 [hep-ph]].

\bibitem{Batell:2011tc}
  B.~Batell, J.~Pradler and M.~Spannowsky,
  JHEP {\bf 1108} (2011) 038
  [arXiv:1105.1781 [hep-ph]].

\bibitem{Kamenik:2011nb}
  J.~F.~Kamenik and J.~Zupan,
  Phys.\ Rev.\ D {\bf 84} (2011) 111502
  [arXiv:1107.0623 [hep-ph]].

\bibitem{Agrawal:2011ze}
  P.~Agrawal, S.~Blanchet, Z.~Chacko and C.~Kilic,
  Phys.\ Rev.\ D {\bf 86} (2012) 055002
  [arXiv:1109.3516 [hep-ph]].

\bibitem{Kumar:2013hfa}
  A.~Kumar and S.~Tulin,
  Phys.\ Rev.\ D {\bf 87} (2013) 9,  095006
  [arXiv:1303.0332 [hep-ph]].


\bibitem{Kile:2013ola}
  J.~Kile,
  Mod.\ Phys.\ Lett.\ A {\bf 28} (2013) 1330031
  [arXiv:1308.0584 [hep-ph]].

\bibitem{Batell:2013zwa}
  B.~Batell, T.~Lin and L.~T.~Wang,
  JHEP {\bf 1401} (2014) 075
  [arXiv:1309.4462 [hep-ph]].


\bibitem{Agrawal:2014aoa}
  P.~Agrawal, M.~Blanke and K.~Gemmler,
  JHEP {\bf 1410} (2014) 72
  [arXiv:1405.6709 [hep-ph]].


\end{thebibliography}
\end{document}